\documentclass[%
reprint,
letter,
 amsmath,amssymb,
 aps,
longbibliography
]{revtex4-1}

\usepackage{graphicx}
\usepackage{dcolumn}
\usepackage{bm}

\begin{document}

\preprint{APS/123-QED}

\title{Plasmon Coupling Induced Photon Scattering Torque}

\author{Yang Li}
\author{Jing Wang}
\author{Hai-Qing Lin}
\affiliation{%
 Beijing Computational Science Research Center, Beijing, 100193, China
}%
\author{Lei Shao}
 \email{shaolei@csrc.ac.cn}
\affiliation{%
 Shenzhen JL Computational Science and Applied Research Institute, Shenzhen, 518131, China
}%
\affiliation{%
 State Key Laboratory of Optoelectronic Materials and Technologies, Sun Yat-sen University, Guangzhou, 510275, China.
}%

\date{\today}

\begin{abstract}
Bio-compatible Au nanoparticles exhibit great advantages in the application of biomedical researches, such as bio-sensing, medical diagnosis, and cancer therapy. Bio-molecules can even be manipulated by laser tweezers with the optically trapped Au nanoparticles as handles. In this Letter, optical scattering torque arising from the coupled Au nanoparticles driven by circularly polarized light is theoretically presented. The coupled plasmon resonance modes boost the angular momentum transfer from photons to the Au nanoparticle dimers and trimers through light scattering, which does not bring any optical-heating side effect. The generated optical torques on the nanostructures highly depend on the plasmon coupling in the structures. The angular momentum transfer efficiencies from scattered photons to nanostructures can reach around 200\%. The results suggest that coupled plasmonic nanoparticle oligomers are promising candidates to construct optically driven rotary nanomotors that can be applied in biomedical applications.
\end{abstract}

\pacs{Valid PACS appear here}
\maketitle


\section{Introduction}
Colloidal Au nanoparticles are promising agents in biomedical researches because of their rich plasmonic properties, well-developed synthetic approaches, chemical inertness, and low cytotoxicity \cite{ELAHI2018537,Storhoff2004,ZhengJP2021}. These nanoparticles therefore are widely used in the applications of surface-enhanced Raman scattering of living cells, clinical diagnosis, biosensing and bioimaging \cite{Kneipp2006,Baptista2007,Bosio2019,Shao2017}. The large optical polarizability of the Au nanoparticles further gives rise to their significantly enhanced optical force when interacting with light, enabling them to be stably trapped by a focused laser beam \cite{Ashkin1987,Lehmuskero2015,Oddershede2015,Hajizadeh2010}. The optical manipulation of Au nanoparticles opens up exciting possibilities for the sophisticated control of bio-molecules, such as clipping DNA or proteins \cite{Wang1997,Heller2013,McCauley2011}.

In addition to enhanced optical forces, large optical torques are enabled by localized surface plasmon resonances supported by the Au nanoparticles because of an efficient angular momentum transfer from photons to the nanostructures. Such angular momentum transfer process, however, is usually accompanied with severe optical heating that will lead to thermal damage of bio-tissues and unstable trapping of the nanoparticles \cite{Lehmuskero2013,Zhou2017,Samadi_2017}. For instance, symmetric spherical Au nanoparticles can be optically driven to rotate fast in an optical tweezers with circular polarization \cite{Lehmuskero2013}. The localized surface plasmon resonances of the Au nanospheres greatly enhance their optical absorption. The spin angular momentum of the absorbed photons are transferred to the nanostructure, producing a remarkably enlarged absorption torque as well as a large plasmonic heating effect.  In contrast to the 100\% absorption-contributed optical torques for the Au nanopheres, asymmetric Au nanorods support both largely enhanced absorption and scattering optical torques, with the scattering contribution to the optical torque reaching 86\% at the localized surface plasmon resonance wavelength \cite{Shao2015}. Besides, multipolar plasmon resonances supported by triangular and square Au nanoplates also boost the angular momentum transfer through scattering \cite{Lee2014}. The scattered photons will transfer angular momentum to the trapped nanostructure but will not cause any photo-thermal effect. The Au nanorods and Au nanoplates with broken rotational symmetry therefore are good candidates in constructing light-driven rotary nanomotors for biomedical applications. However, the preparation of high-quality colloidal Au nanorod and nanoplate samples inevitably introduce the use of the cytotoxic surfactant cetyltrimethylammonium bromide (CTAB), which is difficult to be completely removed \cite{ZhengJP2021,Qin2016,Chen2013,Cano2020,Seaberg2021,Petrilli2018,Maestro2015,Xia2018}.                       

Oligomers of coupled spherical Au nanoparticles with broken rotational symmetry can introduce optical scattering torques as well. The plasmon coupling between the Au nanoparticles results in modulation of the electromagnetic field around the nanostructures, which can control the angular momentum transfer from the photons to the nanostructures \cite{Wu2020}. The dimers/trimers consisting of two or three Au nanospheres in close proximity to each other present rich absorption and scattering spectral features that are interparticle distance-dependent, as well as significantly enhanced electromagnetic near fields in the particle gap \cite{Halas2011}. Such gold nanosphere oligomers perform excellently in various applications, such as nanometric ruler construction \cite{Jain2007}, surface-enhanced Raman scattering \cite{PhysRevLett.83.4357},
and second-harmonic generation \cite{Slablab:12}. Different from the CTAB-capped Au nanorods or nanoplates, Au nanopheres can be synthesized using citrate reduction processes and are stabilized by citrate ligands \cite{citrate_reduction}. The citrate-capped Au nanospheres exhibit reduced toxicity. The loosely bound citrate can also be easily removed by capping agent exchange processes \cite{manson2011polyethylene}. 

In this work, we theoretically investigate the optical torques generated in coupled Au nanoparticles. We calculate the scattering and absorption contributions to the optical torques for Au nanosphere dimers and trimers. Remarkable optical scattering torque can be produced because of the excitation of coupled plasmon resonance modes that strongly depend on the interparticle distance and the conductive connection between the Au nanosphere components in the oligomer structures. The angular momentum transfer efficiencies from scattered photons to nanostructures can reach above 100\%. Our results suggest that coupled plasmonic nanoparticle oligomers are promising candidates as optically driven rotary nanomotors that can be applied in various biomedical applications.

\section{Results}
Both absorption and scattering contribute to the optical torque for nanostructures with broken rotational symmetry, because the scattered photons can have different angular momentum after the scattering event. The total optical torque $\mathbf{M}_{total}=\mathbf{M}_{abs}+\mathbf{M}_{sca}$, with the $\mathbf{M}_{abs}$ and $\mathbf{M}_{sca}$ are the absorption and scattering components. We performed finite-difference time-domain (FDTD) simulations to calculate the electromagnetic field surrounding the coupled Au nanosphere oligomers. The time-averaged $\mathbf{M}_{total}$ is obtained from an area integral of the time-averaged Maxwell Stress Tensor (MST) $\mathbf{T}$ over a closed orientable surface $S$ surrounding the structure
\begin{equation}
\mathbf{M}_{total}=\int(r\times\mathbf{T})\cdot dS.
\label{eq:1}
\end{equation}
The absorption torque is proportional to the numbers of the absorbed photons, and expressed as \cite{Marston1984,Lee2014} 
\begin{equation}
\mathbf{M}_{abs}=\frac{\sigma_{abs}I_{inc}}{\omega},
\label{eq:2}
\end{equation} 
where $\sigma_{abs}$ is the absorption cross section, $I_{inc}$ is the incident light intensity and $\omega$ is the incident field angular frequency. The scattering torque is therefore calculated as
\begin{equation}
\mathbf{M}_{sca}=\mathbf{M}_{total}-\mathbf{M}_{abs}.
\label{eq:3}
\end{equation}

Single Au nanospheres are driven by an optical torque due to the absorption of photons when illuminated by circularly polarized light (CPL) (Fig. 1a). The scattering does not contribute to the optical torque although the scattering cross section is not zero (Fig. 1b), since the photons will not change their polarization state after scattering. When two Au nanospheres are approaching each other and forming a dimer, both absorption and scattering of CPL photons will produce optical torques (Fig. 1c,e,g). The absorption spectra (Fig. 1d,f,h) indicate that a low-energy bonding and a high-energy antibonding hybridized plasmon mode are excited when the dimer is illuminated by a CPL light \cite{Halas2011}. As the gap $D$ between the two Au nanospheres decreases, the bonding dipole plasmon (BDP) mode red shifts dramatically and the antibonding dipole (ABDP) mode blue shifts slightly. The optical torque spectra exhibit similar spectral shape with that of the cross sections. Interestingly, stronger plasmon coupling leads to largely increased contribution from scattering in the optical torque. When the two Au nanospheres are conductively connected by a metallic bridge or the gap between the two Au nanospheres further decreases to a negative value, making the two nanospheres touch each other in a conductive manner, an additional charge transfer plasmon (CTP) resonance mode emerges at the near-infrared region (Fig. 1i--l) \cite{zhu2016quantum}. The CTP resonance of the partially overlapped Au nanosphere dimer exhibits a blue shift compared with the dimer connected by a metallic bridge. Our calculation clearly demonstrates that a non-zero scattering optical torque arises when the Au nanospheres are forming dimers no matter the plasmon coupling is of a capacitive or a conductive manner. The absorption and scattering torques reach their peaks at the wavelength of the hybridized plasmon resonances.

\begin{figure}[ht]
\centering\includegraphics[scale=1]{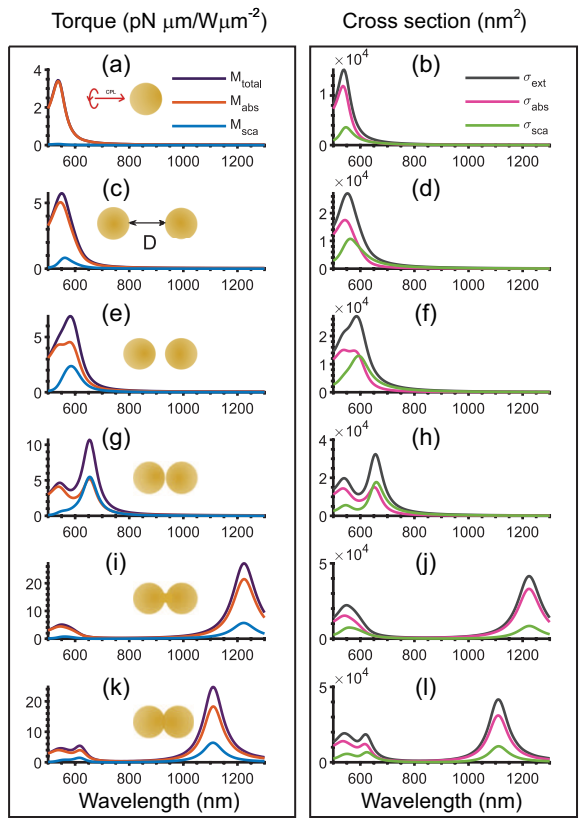}
\caption{Optical torques and cross sections of Au nanosphere dimers consisting of two Au nanospheres with a radius of $30$ nm. From top to bottom, the results of an individual Au nanosphere, nanosphere dimers with a gap distance $D=20$ nm, $D=8$ nm, $D=2$ nm, a nanosphere dimer with the two components seperated by $D=2$ nm and connected by a metallic bridge, and a nanosphere dimer with the two components overlapped with each other with $D=-2$ nm are shown, respectively.}
\label{Fig.1}
\end{figure}

A circularly polarized photon carries an angular momentum of $\pm\hbar$. When the photon is absorbed by the plasmonic nanostructure, all the angular momentum will be transferred to the nanostructure with an transfer efficiency of $100\%$. In contrast, when the plasmonic nanostructure is illuminated by a circularly polarized plane wave, the polarization state of the scattered photons will be determined by the plasmonic near-field. The scattered light may have a spatial phase distribution and thus carry non-zero orbital angular momentum as well \cite{PhysRevLett.110.203906}. The scattering optical torque arising from the angular momentum transfer through scattering therefore can have a different transfer efficiency from the absorption-mediated process. To evaluate the generation efficiency of the scattering optical torque, we plotted the spectra of $M_{sca}/M_{abs}$ for different Au nanosphere dimers, and compared them with the corresponding cross-section ratio $\sigma_{sca}/\sigma_{abs}$ (Fig. 2a,c). $\sigma_{sca}/\sigma_{abs}$ indicates the number ratio between the scattered photons and the absorbed photons by a plasmonic nanostructure and represents the scattering ability of the nanostructure in comparison to its absorption ability. The capacitively coupled Au nanosphere dimer with a gap distance $D=2$ nm exhibits much stronger scattering ability than that of the individual Au nanosphere, while the conductively coupled Au nanosphere dimer with the same gap distance but a metallic bridge presents weaker scattering ability that that of the individual Au nanosphere. Despite the different scattering abilities, both capacitively and conductively coupled Au nanosphere dimers have larger non-zero $M_{sca}/M_{abs}$. Given that the angular momentum transfer efficiency through absorption is always 100\% (see the black dashed lines in Fig. 2b and d), we further calculated the the angular momentum transfer efficiency of the dimer structures through scattering as
\begin{equation}
\eta=\frac{M_{sca}/\sigma_{sca}}{M_{abs}/\sigma_{abs}}=\frac{M_{sca}/M_{abs}}{\sigma_{sca}/\sigma_{abs}}=\frac{M_{sca}/\sigma_{sca}}{N\hbar},
\end{equation}
where $N$ is the absorbed photon number and $\hbar$ is Planck's constant. The angular momentum transfer efficiency $\eta$ is 100\% at 648 nm for the capacitively coupled Au nanosphere dimer, which means each scattered photon transfer 1$\hbar$ angular momentum to the nanostructure. The wavelength is very close to the hybridized BDP mode of the gaped Au nanosphere dimer where its charge distribution is similar to an electric dipole (Fig. 2b). The scattered photons therefore are almost linearly polarized, leading to the 100\% photon angular momentum transfer efficiency. Interestingly, $\eta$ is larger than 100\% in the wavelength range from 591 nm to 648 nm. Similarly, for the conductively coupled Au nanosphere dimer, $\eta=100\%$ at 1155 nm near the CTP resonance and is larger than 100\% in the wavelength range from 833 nm to 1155 nm (Fig. 2d). $\eta$ reaches its peak of 200\% at 926 nm for such bridged Au nanosphere dimer. Further analysis revealed that the bridged Au nanosphere dimer exhibits a weak transverse dipole-dipole antibonding mode at 575 nm \cite{Marinica2012} and a high-order hybridized plasmon mode at 926 nm. The excitation of the 575-nm mode results in an imperfect dipole-like charge distribution on the surface of the bridged dimer and $\eta=50\%$. At the CTP resonance wavelength, the charge distribution on the dimer is almost a perfect electric dipole \cite{Wen2015}, leading to $\eta=100\%$. Surprisingly, the excitation of the high-order hybridized mode gives rise to $\eta=200\%$ because of the specific surface charge distribution as well as the resulted electromagnetic near field. 

\begin{figure}[htbp]
\centering\includegraphics[scale=1]{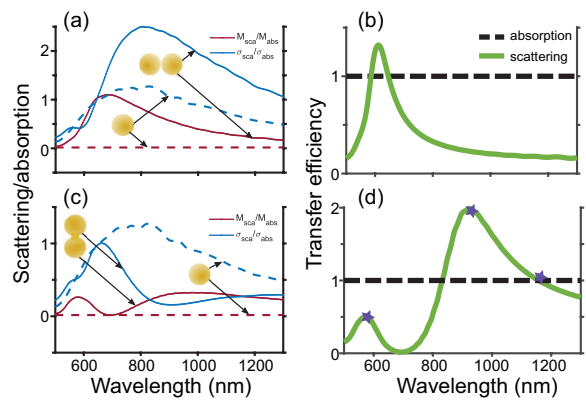}
\caption{Scattering abilities and scattering angular momentum transfer efficiencies of Au nanosphere dimers. (a) Spectra of $M_{sca}/M_{abs}$ and $\sigma_{sca}/\sigma_{abs}$ for a capacitively coupled Au nanosphere dimer with a gap of 2 nm in comparison with those of an individual Au nanosphere. (b) Angular momentum transfer efficiencies for scattering and absorption of the Au nanosphere dimer in (a).(c) Spectra of $M_{sca}/M_{abs}$ and $\sigma_{sca}/\sigma_{abs}$ for a conductively coupled Au nanosphere dimer with a gap of 2 nm and a metallic bridge in comparison with those of an individual Au nanosphere. (d) Angular momentum transfer efficiencies for scattering and absorption of the Au nanosphere dimer in (c).}
\label{Fig.2}
\end{figure}

Our results show that the angular momentum transfer efficiency of the scattered photons by the Au nanosphere dimers, no matter capacitively coupled or conductively coupled, can be larger than 100\%. The rod-shaped Au nanoparticle has been reported to exhibit the same behavior, with the angular momentum transfer efficiency through scattering larger than 100\% at the wavelength range between the rod transverse plasmon resonance and longitudinal plasmon resonance \cite{Shao2015}. The individual nanoparticle can be described as a dipole in the quasistatic approximation with its polarizability $\text{\ensuremath{\alpha(\omega)=\alpha'(\omega)+i\alpha''(\omega)}}$, where $\omega$ is the incident light frequency \cite{bohren2008absorption,zeman1987accurate}. An elongated Au nanoparticle can be modeled by an ellipsoid with its polarizability being analytically expressed. The nanoparticle dipole moment therefore is
\begin{equation}
\mathbf{p}=\varepsilon_{0}\varepsilon_{m}\alpha(\omega)\mathbf{E},
\end{equation}
where $\varepsilon_{0}$ is the vacuum permittivity and $\varepsilon_{m}$ is relative permittivity of the surrounding medium.  We have employed such model to calculate the plasmonic dipole moment and the cross sections of a prolate Au nanospheroid with its semi-major axis of 60 nm and semi-minor axis of 30 nm illuminated by a right-handed circularly polarized (RCP) light (Fig. 3). The calculated cross-section spectra can well reproduce the results obtained from numerical simulation (Fig. 3a). Two resonance peaks are observed at 525 nm and 640 nm, clearly indicating the transverse and longitudinal localized surface plasmon resonances that correspond to the electron oscillations along the minor and major axis of the spheroid, respectively. The phase difference between the two in-plane perpendicular components of the plasmonic dipole moment reaches above 0 at some wavelengths in between the two resonance wavelengths, i.e. 525 nm and 640 nm (Fig. 3b). Given that the phase difference for the incident RCP light is $-0.5\pi$, some of the photons change their polarization state from right-handed circularly polarized to left-handed circularly polarized after scattering. The angular momentum change of such photons therefore is $2\hbar$ after scattering. As a result, the angular momentum transfer efficiency through scattering $\eta$ can be larger than 100\% at the above wavelengths (Fig. 3c). Base on the discussions above, we believe the same mechanism dominates in the angular momentum transfer in Au nanosphere dimers, i.e. it is the change of the phase difference between the two components of the excited plasmonic dipole moment of the whole structure that mainly contributes to $\eta>100\%$ at the shorter wavelength side of the BDP or CTP resonances of the dimers in Fig. 2b and d.

\begin{figure}[t]
\centering\includegraphics[scale=1]{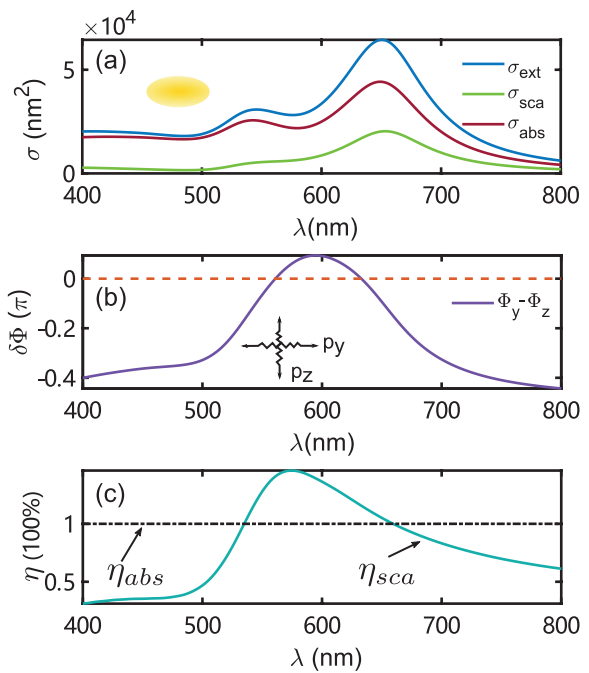}
\caption{Dipole model for the analysis of angular momentum transfer process. A prolate Au ellipsoid with major axis of 120 nm and minor axis of 60 nm is analyzed. (a) Optical cross sections calculated for the dipole model. (b) Phase difference between the two in-plane perpendicular components of the plasmonic dipole moment. (c) Angular momentum transfer efficiencies of the absorption and scattering calculated by FDTD simulation.}
\label{Fig.3}
\end{figure}

In addition to the Au nanosphere dimers, we further studied the optical torques in the Au nanosphere trimers consisting of three Au nanospheres with the same radius of 84 nm. The smallest gap distances between the nanospheres were set at 2 nm. When illuminated by circularly polarized light, the nanosphere trimer has a large scattering cross section at the NIR wavelengths (Fig. 4a,b). However, the scattering torque is small at the scattering cross section peak. Instead, the trimer exhibits a strong scattering optical torque peak at around $\lambda=775$ nm (Fig. 4c). The corresponding scattering angular momentum transfer efficiency reaches its peak at the same wavelength (Fig. 4d), with the peak value of 150\%. To ascertain the origin of the high angular momentum transfer efficiency, we calculated the azimuthal phase distribution after the light is scattered by the trimer (Fig. 4e). The phase distribution result suggests that the Au sphere trimer generates a non-zero orbital angular momentum (OAM) in the near field of the plasmonic nanostructure. Calculated charge distribution of the trimer structure shows that a high order plasmon mode appears at the peak wavelength of the optical torque spectrum (Fig. 4f). The excitation of the subradiant high order plasmon hybridized resonance mode leads to the larger scattering optical torque, in agreement with the results reported for triangular and square Au nanoplates \cite{Lee2014}.

\begin{figure}
\centering\includegraphics[scale=1]{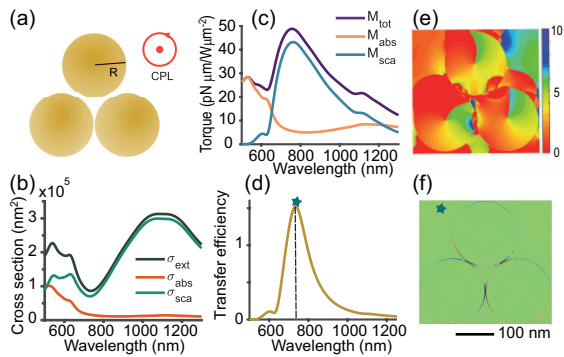}
\caption{Optical torques and cross sections of a trimer of Au nanospheres with the sphere radius of $R=84$ nm and the interparticle gap distance of $2$ nm. (a)Schematic showing the Au nanosphere trimer. (b) Optical cross sections of the Au nanosphere trimer. (c) Optical torques of the Au nanosphere trimer. (d) Angular momentum transfer efficiency of the scattering. (e) Calculated azimuthal phase distribution after the light is scattered by the trimer. (f) Charge distribution of the central cross section of the Au nanosphere timer at the peak wavelength of the optical torque spectrum.}
\label{Fig.4}
\end{figure}

In summary, we studied the optical torques of Au nanosphere dimers and trimers when excited by circularly polarized light. We found that the coupled plasmon resonances in the Au nanosphere oligomers can boost the angular momentum transfer from photons to the nanostructures through scattering, with the transfer efficiency reaching as high as 200\%. The phase difference between different components of the plasmonic dipole and the excitation of the high-order hybridized plasmon resonance modes are the main reasons for the high angular momentum transfer efficiencies. The generation of scattering-dominant optical torque is highly demanded because it can minimize the side effect brought by plasmonic photothermal heating. Our results suggest that coupled plasmonic nanoparticle oligomers are promising for constructing optically driven rotary nanomotors that have potential applications in areas like biosensing, nanosurgery, and biomedical therapies.

\begin{acknowledgments}
This work was financially supported by National Natural Science Foundation of China (Grant No. NSAF U1930402) and State Key Laboratory of Optoelectronic Materials and Technologies of China (Grant No. OEMT-2019-KF-07). L.S. acknowledges the support from the Pearl River Talent Recruitment Program (2019QN01C216) and computational resources from the Beijing Computational Science Research Center.
\end{acknowledgments}

\nocite{*}

\bibliography{m20211117B}

\end{document}